\documentclass[online]{aa}

\usepackage{graphicx}
\usepackage{txfonts}
\usepackage{natbib}
\usepackage{color}
\usepackage{breakurl}
\usepackage{enumitem}
\usepackage{siunitx}
\usepackage{hyperref}
\hypersetup{colorlinks,citecolor=blue,urlcolor=blue,linkcolor=blue}
\usepackage{amsmath}
\usepackage{xcolor}

\usepackage[section]{placeins}
\usepackage{placeins}
\usepackage{float}
\usepackage{xcolor}
\let\Oldsection\section
\renewcommand{\section}{\FloatBarrier\Oldsection}
\let\Oldsubsection\subsection
\renewcommand{\subsection}{\FloatBarrier\Oldsubsection}
\let\Oldsubsubsection\subsubsection
\renewcommand{\subsubsection}{\FloatBarrier\Oldsubsubsection}

\newcommand{\orcid}[1]{\href{https://orcid.org/#1}{\includegraphics[width=10pt]{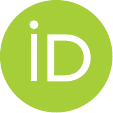}}}

\newcommand{\doiurl}[1]{\href{https://doi.org/#1}{#1}}

\graphicspath{{.}{Plots/}}

\begin{document} 

\setlength{\topmargin}{-20pt}
\AANum{A184}
\yearCop{2023}
\doi{\doiurl{10.1051/0004-6361/202346312}}
\idline{A\&A 678, A184 (2023)}
\hypersetup{
	pdftitle = {Coronal bright point statistics \\ I. Lifetime, shape, and coronal co-rotation},
	pdfauthor = {I.~Kraus, Ph.-A.~Bourdin, J.~Zender, M.~Bergmann, A.~Hanslmeier},
	pdfkeywords = {Sun: corona -- Sun: UV radiation -- Methods: observational -- Methods: statistical},
	pdfsubject = {Astronomy \& Astrophysics}
}

\title{Coronal bright point statistics --- I. Lifetime, shape, and coronal co-rotation}
\titlerunning{Coronal bright point statistics ~~ I. Lifetime, shape, and coronal co-rotation}
\authorrunning{I. Kraus et al.}

\author{I. Kraus \inst{1} \orcid{0000-0002-0451-811X},
          Ph.-A. Bourdin \inst{1, 2} \orcid{0000-0002-6793-601X},
          J. Zender \inst{3} \orcid{0000-0003-2728-3664},
          M. Bergmann \inst{3, 4},
          A. Hanslmeier \inst{1}}

\institute{Institute of Physics, University of Graz, Universit{\"a}tsplatz 5, 8010 Graz, Austria\\
              \email{Isabella.Kraus@uni-graz.at, Philippe.Bourdin@uni-graz.at}
         \and
             Space Research Institute, Austrian Academy of Sciences, Graz, Austria
         \and
             European Space Research and Technology Center, 2200 AG Noordwijk, Netherlands
          \and
             Iceye Oy, Maarintie 6, 02150 Espoo, Finland}

\date{Received 3 March 2023 / Accepted 18 July 2023 / Published 20 October 2023}


\abstract
{The corona of the Sun is the part of the solar atmosphere with temperatures of over one million Kelvin, which needs to be heated internally in order to exist.
This heating mechanism remains a mystery; we see large magnetically active regions in the photosphere lead to strong extreme UV (EUV) emission in the corona. On much smaller scales (on the order of tens of $\unit{Mm}$), there are bipolar and multipolar regions that can be associated with evenly sized coronal bright points (CBPs).}
{Our aim was to study the properties of CBPs in a statistical sense and to use continuous data from the SDO spacecraft, which makes it possible to track CBPs over their whole lifetime.
Furthermore, we tested various rotation-speed profiles for CBPs in order to find out if the lower corona is co-rotating with the photosphere.
Then we compiled a database with about 346 CBPs together with information of their sizes, shapes, appearance and disappearance, and their visibility in the EUV channels of the AIA instrument. We want to verify our methods with similar previous studies.}
{We used the high-cadence data of the largest continuous SDO observation interval in 2015 to employ an automated tracking algorithm for CBPs.
Some of the information (e.g., the total lifetime, the characteristic shape, and the magnetic polarities below the CBPs) still requires human interaction.}
{In this work we present statistics on fundamental properties of CBPs along with some comparison tables that relate, for example, the CBP lifetime with their shape.
CBPs that are visible in all AIA channels simultaneously seem to be brighter in total and also have a stronger heating, and hence a higher total radiation flux.
We compared the EUV emission visibility in different AIA channels with the CBP's shape and lifetime.
From the tracking algorithm we confirm a strict co-rotation of the CBPs with the photospheric differential rotation.}
{The tracked CBPs have a typical lifetime of about 1--6 hr, while the hottest and brightest ones seem to exist for significantly longer time, up to 24 hr.
Furthermore, the merging of two CBPs seems not to have an influence on the overall size of the persisting CBP.
Finally, fainter and cooler CBPs tend to have only weaker magnetic polarities, which clearly supports a coronal bright point heating mechanism based on magnetic energy dissipation.}


\maketitle

%
\section{Introduction}

Coronal bright points (CBPs) are hot plasma phenomena that are located in the lower corona \citep{2014AN....335.1037K,2014AstL...40..510M}.
CBPs were first observed in X-rays \citep{1973SoPh...32...81V}.
These phenomena were interpreted as small EUV-bright structures in the solar atmosphere \citep{1979SoPh...63..119S}.

Temperatures of CBPs could be above $1\ \unit{MK}$ and up to $3.4\ \unit{MK}$ \citep{2011A&A...526A..78K}.
A typical CBP has a life span that ranges from minutes to days, and a CBP is usually smaller than 60 arcsec in diameter \citep{1973ApJ...185L..47V,2001SoPh..198..347Z}.
Previous works showed that CBPs could be formed from the reconnection of bipolar magnetic fields \citep{1976SoPh...50..311G,2007PASJ...59S.735K,2012Ap&SS.341..215L}.
CBPs seem to appear preferably for horizontal background flux density of $3-6\ \unit{G}$ \citep{2001ApJ...553..429L}.
Magnetic fluxes of the polarities are in the range of about $10^{19}-10^{20}\ \unit{Mx}$ \citep{1976SoPh...50..311G}.
For a statistical CBP study \cite{2003ApJ...589.1062H} introduced a CBP selection criterion that is not only based on spatial, temporal, and magnetic properties, but also based on the CBP shape, in particular on its eccentricity.
It is possible to characterize CBP shapes as point-like, loop-like, and multiple-loop complex structures.
The number of CBPs seems to be independent of the solar cycle \citep{2002ApJ...564.1042S,2003ApJ...589.1062H,2011A&A...526A..78K}.

Short reconnection events with limited spatial extent may provide a minor contribution to the total coronal energy.
However, if we multiply by the number of CBPs on a quiet solar disk, which means by a factor of about 1000, this magnetic energy release has a somewhat larger contribution for the heating of the corona  because CBPs are found in a relatively uniform distribution \citep{2014AN....335.1037K} with up to 1500 CBPs per day in  quiet-Sun and coronal hole regions \citep{1974ApJ...189L..93G,1976SoPh...49...79G}.
CBPs cover about 1.4\% of the quiet-Sun area but contribute to about 5\% of the quiet-Sun radiation \citep{2001SoPh..198..347Z}.
All CBP energy releases together are estimated to about $2 \cdot 10^{22}\ \unit{J}$, and CBPs can usually flare more than once in their lifetime \citep{2014AN....335.1037K,2014AstL...40..510M,2014ApJ...784..134R}.
Given these numbers, we compute that CBPs contribute with an average of $45\ \unit{W/m^2}$ to the quiet-Sun coronal heating, which is 4.5\% of the total coronal required energy input.
A comprehensive review of CBP properties and statistics can be found in \cite{2019LRSP...16....2M}.

In this work our aim is to confirm our CBP detection and tracking methods. To this end, we compare our results with previous similar studies. In a next step, this will allow further studies on the magnetic polarities below CBPs. In Sect.~\ref{S-methods} we explain our methods, and in Sect.~\ref{S-results} we show statistical properties of our CBP ensemble.

\section{Methods}\label{S-methods}

We searched for the longest and continuous data set without any missing data and the least active regions in the year 2015.
Our selected observation period therefore spans 12 days: 13--24 August 2015.

The extension of the data set of tracked CBPs over a period of more than 12 days was not possible due to larger gaps in the available data.
There was an Earth eclipse from 25 August 2015 to 9 September 2015, due to the circular geosynchronous orbit of Solar Dynamics Observatory (SDO) around Earth.
The other reason was an orbit maintenance maneuver on 12 August 2015.
In general, further interruptions are possible, such as Earth eclipses; Moon, Venus, and Mercury transits; and several spacecraft maneuvers.

We used SDO level 1 data with $4096 \times 4096$ pixels and a resolution of 0.6 arcsec per pixel \citep{2012SoPh..275....3P,2012SoPh..275...17L,2012SoPh..275...41B}.
We used ``aia\_prep'' from the {\em{SolarSoft}} package to process the images into level 1.5 data products.
The Atmospheric Imaging Assembly (AIA) images are co-rotated and co-aligned with the orientation of the Helioseismic and Magnetic Imager (HMI) level 1.5 images \citep{2012SoPh..275..207S,2012SoPh..275..229S}.

\subsection{Solar disk segmentation}
The segmentation of the solar disk into active regions (ARs), coronal holes (CHs), and the quiet-Sun (QS) was performed using the {\em{SPoCA}} algorithm \citep{2014A&A...561A..29V}, and is described in detail by \cite{2014A&A...561A...9K}.
The algorithm uses AIA (17.1\ \unit{nm}, 19.3\ \unit{nm}) images and was improved later on \citep{2017A&A...605A..41Z} to extend the AR area taking information from the AIA magnetograms, provided by the HMI instrument, into account.
This enlargement of the ARs ensures that the region is not underestimated.
The CBP identification is based on image morphological operators \citep{Haralick1987} applied to AIA 193\ \unit{\AA} images with a resolution of $4096 \times 4096$ pixels.
Further details of the algorithm are described in \cite{2021SoPh..296..138V}.

\subsection{Tracking CBPs}\label{S:2_3_6_Tracking_and_creating_videos}

Coronal bright points can automatically be identified via different selection criteria that depend on the CBP size \citep{2014ApJ...784..134R,2010SoPh..262..321S}, shape \citep{2002A&A...392..329B,1979SoPh...63..119S,2015ApJ...807..175A}, proximity to ARs \citep{2015ApJ...807..175A}, and intensity \citep{2010A&A...516A..50S,2012A&A...538A..50S}. 

We describe here the automatic tracking algorithm we implemented based on the following selection criteria:
\begin{enumerate}[label=\arabic*)]
\item a CBP is brighter than a given threshold in the normalized intensity of the AIA 171 channel, where the threshold is above the background noise level and defines the onset of a CBP;
\item the candidate CBP fits to an ellipse with an eccentricity of less than 2.5 to avoid confusion with coronal loops or other strongly elongated structures;
\item the CBP is larger than 200 and smaller than 2500 AIA pixels to avoid tracking small flares (lower limit) or a full active region (upper limit);
\item the width and height are both smaller than 30 arcsec to avoid tracking brightenings due to reconnection in a complex magnetic network;
\item CBPs are at least 180 arcsec away from the center of active regions (ARs) to avoid as many plage regions as possible around ARs;
\item CBPs have to be at least 5\% of the solar radius away from the limb, due to a large geometrical distortion near the limb;
\item the lifetime of a CBP is at least 60 minutes because it is important to study the onset, the evolution, and the disappearance of CBPs, as well as the rotation behavior of the corona, which requires a sufficiently long tracking of each CBP;
\item the CBP area never changes by more than 30\% between two consecutive images, which is expected for regular CBPs because they do not change significantly within one minute.
It should be noted that the last criterion was never triggered for any of our CBPs.\end{enumerate}

When we tracked CBPs over a longer time, we needed to take into account the latitude-dependent differential rotation of the Sun.
Since it is not known if the lower corona really co-rotates with the photosphere, we tried out different differential rotation profiles for the lower corona. 

Our first approach was the pixel coordinate method, where the tracking algorithm finds the center of mass in the vicinity of the CBP position. Then we tracked a CBP from image to image. To do this we place a box around the CBP and the algorithm finds the new centroid. The shift speed of the box is not the same for all CBPs, due to differential rotation. Without this method, the centroid shifts out of the box after 50 minutes. The exact speed of the CBP is not known in advance, and therefore it is not possible to move the box at a pre-defined speed. After ten minutes the box already begins losing the centroid.

In a second approach, we extended the algorithm by a pre-defined solar rotation profile that depends on the latitude and corrects the CBP position accordingly. We tried different rotation profiles, including a strict co-rotation with the photosphere.
The change from one image to the next was negligible. This method uses a rotating heliographic coordinate system, which rotates exactly like the differential rotation in the photosphere. When a CBP rotates at the same speed as the heliographic coordinate grid, it has approximately the same latitude and longitude over its lifetime.

If this differential rotation profile does not exactly fit the proper motion of the CBPs on the solar disk, the tracking will eventually fail.
The reason is that the CBP can wander off the field of view after a few hours, which is significantly shorter than most of the CBP lifetimes.
This allowed us to test the CBP rotation with respect to the photospheric rotation.
In the following, we use only the photospheric differential rotation profile for the CBP location tracking, which is the only method that works for our analysis.

\begin{figure}[!htbp]
\includegraphics[width=9cm]{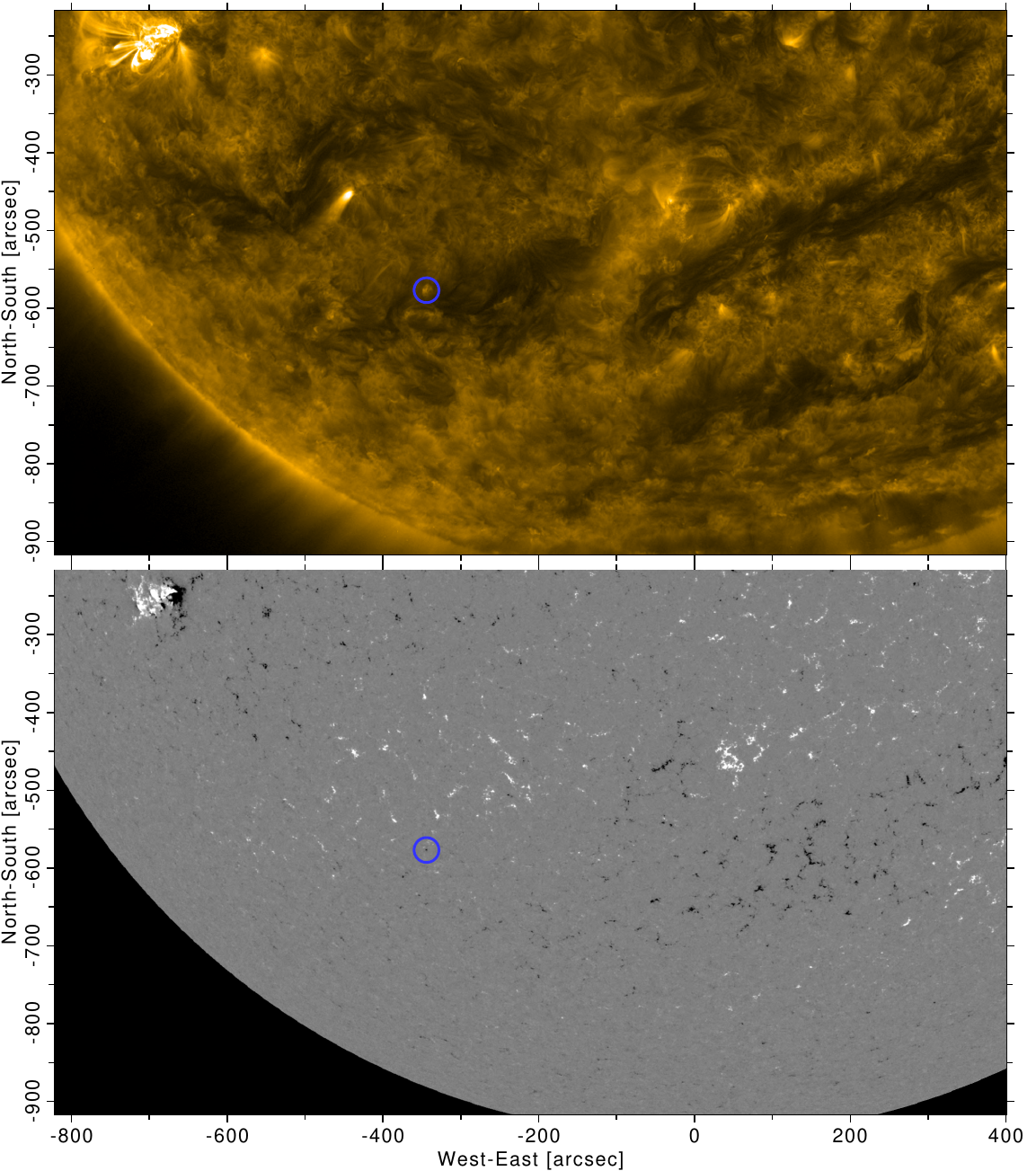}
\caption{Example of a CBP (blue circle) as seen in the 171 \unit{\AA} channel in the middle of its lifetime. The image was taken on 14 August 2015 at 10:11:09 UT, which is same as the CBP in the left panel of Fig.~\ref{F:Overview_Shapes}. The coordinates are given as helioprojective, solar north is up. A video of this CBP's evolution is available online (see \href{https://doi.org/10.5281/zenodo.6610809}{DOI:~10.5281/zenodo.6610809}).}
\label{F:AIA171_Snapshot_Contours_Ausschnitt_Kreis}
\end{figure}

Each of the AIA frames is processed separately and all CBPs are marked with contour lines (see Fig.~\ref{F:AIA171_Snapshot_Contours_Ausschnitt_Kreis}).
In the next step, the AIA frames are stacked with respect to the differential rotation.
If a CBP reappears at the same location, it inherits a unique ID number from the previous frame.
Therefore, each CBP can be tracked throughout its whole lifecycle.
In the final step, the classification of the shapes of the CBP is done manually.

To identify the appearance and disappearance of a CBP, we analyze the normalized total EUV intensity.
When the intensity reaches a specific value higher than a given threshold plus one standard deviation of this intensity, we interpret it as the appearance of a new CBP.
If the intensity drops below this value, we note this as the disappearance of the CBP.
This results in some flickering of the tracking with the appearance and disappearance of the CBPs.
To compute the lifetime of each CBP, we determine the actual begin and end times by human inspection.

\subsection{EUV flux evolution}\label{S-intensityevolution}

\begin{figure}[!htbp]
\centering
\includegraphics[width=8.5cm]{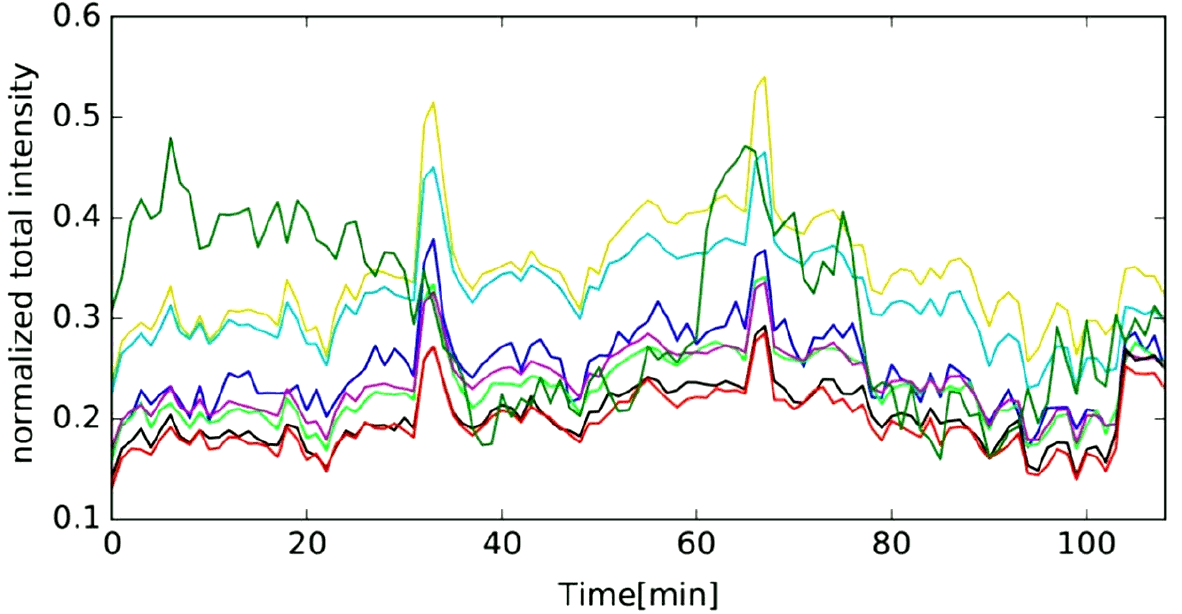}
\caption{Evolution of normalized EUV flux of one example CBP.
Shown are the AIA channels 94 \unit{\AA} (blue), 131 \unit{\AA} (light green), 171 \unit{\AA} (yellow), 193 \unit{\AA} (cyan), 211 \unit{\AA} (magenta), 304 \unit{\AA} (black), and 335 \unit{\AA} (red).
The dark green line indicates the total unsigned flux from HMI (see Sect.~\ref{S-intensityevolution}).}
\label{F:Figure_3_3_BP_intensities_images45_bp165}
\end{figure}

In Figure~\ref{F:AIA171_Snapshot_Contours_Ausschnitt_Kreis} we show the AIA 171 total intensity of one example CBP near the middle of its lifecycle.
We further investigate the CBP identified with a blue circle, and plot its temporal evolution of the EUV flux in different AIA channels (see Fig.~\ref{F:Figure_3_3_BP_intensities_images45_bp165}).
We find that during the lifecycle of the CBP the EUV flux rises consistently across different wavelengths, which roughly indicates the beginning and end of the lifecycle.

We normalized all our CBP examples with the maximum EUV flux of one randomly selected average CBP as reference in order to make the EUV fluxes of all CBPs comparable with each other.
For one case study of the main phase of a CBP, we find short-lived and high peaks in the EUV flux, seen at minutes 33 and 67 in Fig.~\ref{F:Figure_3_3_BP_intensities_images45_bp165}.
The EUV emission peak at minute 33 follows a significant decrease in the total unsigned HMI flux, which may indicate that this magnetic energy is dissipated during the appearance phase of the CBP.
The second EUV peak at minute 67 follows a very significant and long-lasting increase in the total unsigned HMI flux; this increase started about seven minutes prior to the EUV peak.
This time delay between the rise in the HMI flux and the strong brightness enhancements of the EUV intensity in all the AIA channels roughly coincides with the Alfv{\'e}n travel time from the photosphere to the lower corona.

The reason for the appearance of a CBP is most probably due to a magnetic energy release during reconnection processes \citep{2003A&A...398..775M}.
The higher the photospheric magnetic flux, the more energy generated in the lower corona, which consequently leads to stronger EUV emission or larger CBPs \citep{2013SoPh..286..125C}.

\subsection{Data selection for analysis} \label{S-dataselection}

Following our general criteria, we are able to track 663 CBPs.
Due to CBP tracking misses during the appearance or the disappearance phase of CBPs, not all CBPs are covered for their whole lifetime.
In the following we restrict our analysis to only those 346 CBPs (52\%) of the whole ensemble, which are tracked completely from their appearance to their disappearance.
Our lifetime statistics therefore contains data only from those complete CBP tracking sequences.
Since our ensemble now contains
\begin{math}
{N=346}
\end{math}
CBPs, the statistical error amounts to about
\begin{math}
{\sqrt{N}/N = \pm 5.4\%.}
\end{math}

\section{Results}\label{S-results}

\subsection{Visibility}\label{S-visibility}

\begin{figure}
\centerline{\includegraphics[width=8.8cm]{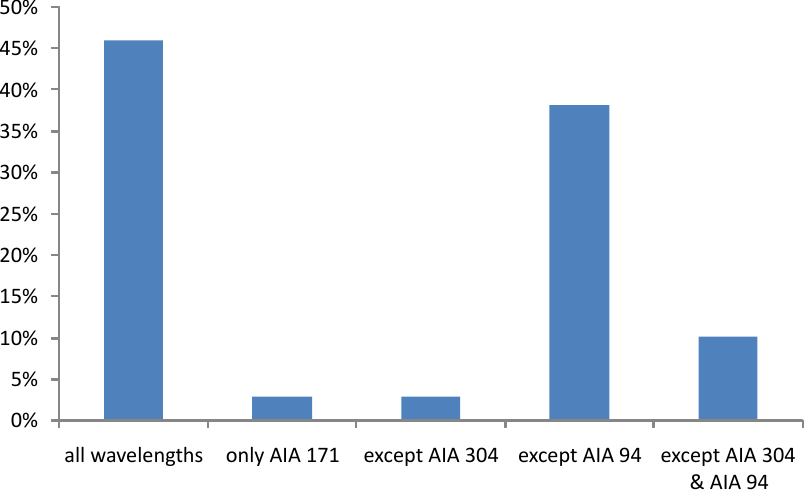}}
\caption{Visibility of CBPs in different combinations of AIA channels (see Sect.~\ref{S-visibility}).}
\label{F:5_Temperature}
\end{figure}

In this section we discuss in which AIA channel CBPs are visible.
The channels roughly correspond to different ranges in the coronal plasma temperature \citep{2012SoPh..275...17L,2012SoPh..275...41B}.
We find that 46\% of the CBPs are visible in all AIA channels (see Fig.~\ref{F:5_Temperature}).
CBPs that are visible in all channels are usually also the brightest ones, and therefore probably have a high coronal temperature near and above $1$ MK.
The second-largest category is visible in all channels except for the AIA 94 channel, which is generally the AIA channel with the least intensity and greatest instrumental noise.
If CBPs are visible in fewer channels, they have a tendency to be fainter overall, and hence probably cooler, and perhaps located at lower height.
As few as 3\% of the CBPs are visible only in the AIA 171 channel, which makes these CBPs the faintest ones.

In Table~\ref{T:Table_3_5_1_Temperature_probability_channel} we show the fraction of CBP visibility in the different AIA channels.
We used the AIA 171 channel to select CBPs because the AIA 171 response function supersedes all other channels in the expected temperature range below 1.25 MK for forming and fainter CBPs. Hotter and hence brighter CBPs that become visible in the AIA 193 channel are always visible in the AIA 171 channel. Therefore, 171 was the right channel to base our selection criterion on.
We see that CBPs are simultaneously observable in AIA 171, 131, 193, and 211 in more than 90\% of all cases (see Table~\ref{T:Table_3_5_1_Temperature_probability_channel}).
This is not unexpected because these AIA channels have some overlapping intervals in their temperature response functions.
We find about 84\% of the CBPs are visible in the AIA 304 channel.
This means there is also emission from cooler plasma associated with the CBP, even though we cannot say for sure from what height this cooler helium emission originates.
A large fraction of CPBs (91\%) are visible in at least four AIA channels (171, 131, 193, and 211) and 84\% are visible in at least five AIA channels (171, 131, 193, 211, and 304); we thus find that mostly all our CBP trackings are significantly above the noise level of the AIA instrument (see Table~\ref{T:Table_3_5_1_Temperature_probability_channel}).
Even in the AIA 335 channel 63\% of our CBPs are visible, which is again a good indication that we observe most of the CBPs significantly above the noise level even in this channel with a relatively low response function.
For the AIA 94 channel we know this channel also has a low response function and is sensitive mainly to very hot plasma.
Therefore, we do not expect to see many CBPs in this channel.
Still, about half of all our tracked CBPs are visible in the AIA 94 channel. In the following sections we use the observability in AIA channels with the classes ``all channels'' (bright CBPs), ``more than one but not all channels'' (less bright CBPs), and ``only AIA 171'' (faint CBPs).

\begin{table}[!htbp]
\centering
\caption{Fraction of CBP visibility in the different AIA channels.}
\label{T:Table_3_5_1_Temperature_probability_channel}
\begin{tabular}{cc}
\hline \hline
AIA channel & CBPs visibility \\
\hline
304 \unit{\AA} &  84\%\\
131 \unit{\AA} &  95\%\\
171 \unit{\AA} & selection criterion\\
193 \unit{\AA} &  94\%\\
211 \unit{\AA} &  91\%\\
335 \unit{\AA} &  63\%\\
94 \unit{\AA}  &  49\%\\
\hline
\end{tabular}
\end{table}

\subsection{Shape}\label{S-shape}

\begin{figure}[!htbp]
\centering
\includegraphics[width=8.8cm]{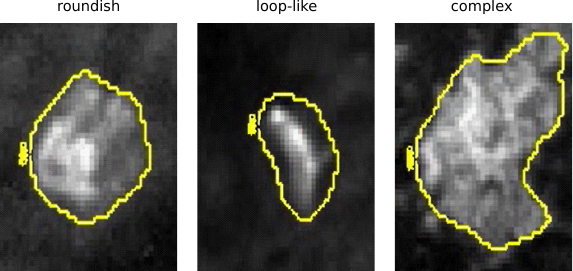}
\caption{Overview of CBP shape classes. From left to right: round (\href{https://doi.org/10.5281/zenodo.6610809}{DOI:~10.5281/zenodo.6610809}), loop-like (\href{https://doi.org/10.5281/zenodo.6610821}{DOI:~10.5281/zenodo.6610821}), and complex (\href{https://doi.org/10.5281/zenodo.6610842}{DOI:~10.5281/zenodo.6610842}). Each panel has an apparent size of about $30 \times 42~\unit{arcsec}$ or $50 \times 70$ SDO pixels.}
\label{F:Overview_Shapes}
\end{figure}

We  also wanted to investigate the spatial shape of CBPs, which we compare later to other properties of CBPs (see Sects.~\ref{S-Visibility-shape} and ~\ref{S-lifetimevsshape}).
We find that 49\% of the CBPs have a loop-like structure (see Fig.~\ref{F:Overview_Shapes}).
A minority of about 16\% is mostly round and of simple geometry;
 35\% of the CBPs have complex shapes.

In the online material we show sample videos of the tracking for a round CBP (2015-08-14 07:38:09 UT, \href{https://doi.org/10.5281/zenodo.6610809}{DOI:~10.5281/zenodo.6610809}), a loop-like CBP (2015-08-24 03:56:08 UT, \href{https://doi.org/10.5281/zenodo.6610821}{DOI:~10.5281/zenodo.6610821}), and a CPB with a complex shape (2015-08-14 12:15:39 UT, \href{https://doi.org/10.5281/zenodo.6610842}{DOI:~10.5281/zenodo.6610842}).


\subsection{Appearance}\label{S-appearance}

We now analyze the onset phase of the EUV emission of CBPs while they appear.
For about 28\% of the CBPs this onset phase is unclear because of too many fluctuations in the EUV emission, which is the reason why we leave these unclear CBP appearances out of our further discussion.
We divide the EUV emission onset duration into three classes: less than 5 min (fast), 5--10 min (medium), 10 min or longer (slow).
Surprisingly, CBPs seem to either appear fast or slow (see Fig.~\ref{F:fastslow_appearance_a_disappearance_complete}).
This means that we find a bi-modal distribution of the onset duration.

In Table~\ref{T:Table_3_9_1_1_Pre-Interaction_properties} we describe the preconditions and the appearance phase of the CBPs, where multiple options are possible.
For about 60\% of the appearing CBPs we observe a slightly fluctuating EUV flux.
Unfortunately, this leads to some tracking gaps in the appearance phase of CBPs because the algorithm uses a fixed threshold in the AIA 171 channel to identify a CBP.
Therefore, we need to make a final human control to close these tracking gaps.

We confirm previous studies that also find short-lived and strong variations in the light intensity \citep{2010ASSP...19..424P,2014AN....335.1037K}.
This supports the notion that fast and spatially limited magnetic energy dissipation may contribute to an impulsive heating of CBPs.

We also find that 40\% of the CBPs grow while they appear, 21\% separate from another pre-existing CBP, and 14\% merge with another CBP.
For less than 4\% of the CBPs we observe by visual human inspection that a CBP changes its shape category during their appearance (e.g. from round to complex shape; see the appearance of a complex CBP) (\href{https://doi.org/10.5281/zenodo.6610842}{DOI:~10.5281/zenodo.6610842}).

\begin{table}[!htpb]
\centering
\caption{CBP emergence options, multiple choices possible.}
\label{T:Table_3_9_1_1_Pre-Interaction_properties}
\begin{tabular}{lr}
\hline \hline
Properties of emergence & Probability\\
\hline
appears after another CBP disappeared   &   1.4\% \\
separating from another CBP             &  20.5\% \\
merging with another CBP                &  13.6\% \\
fluctuating intensity                   &  59.5\% \\
area shrinks                            &   0.3\% \\
area grows                              &  39.6\% \\
shape changes                           &   3.5\% \\
unclear                                 &   0.3\% \\
\hline
\end{tabular}
\end{table}

\begin{figure}[!htbp]
\centering
\includegraphics[width=7cm]{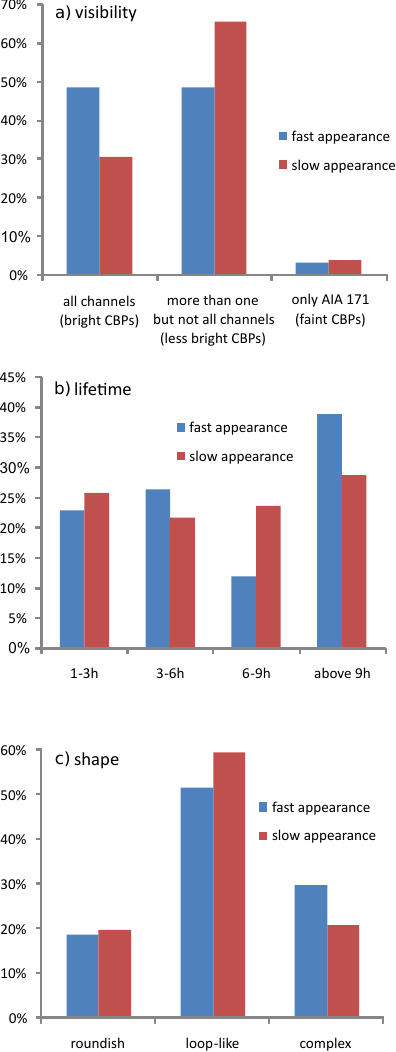}
\caption{Separation of fast-appearing (blue) and slow-appearing (red) CBPs vs. the visibility (upper panel), the lifetime (middle panel), and the shape (lower panel) (see Sect.~\ref{S-appearance}).}
\label{F:9_2_fastslow_appearance}
\end{figure}

We now discuss the relation the between fast and slow appearance of CBPs with respect to other properties, such as their visibility in the AIA channels, their lifetime, and their shape.
From the upper row in Fig.~\ref{F:9_2_fastslow_appearance} we learn that CBPs are visible in many AIA channels, which means they are usually brighter in general, are also appear more quickly.
In the opposite case, less bright CBPs are visible in fewer channels and tend to appear more slowly.

Furthermore, we find that slow-appearing CBPs have a significantly smaller fraction of lifetimes longer than 9 hr, but the fraction of CBPs living for 6--9 hr increases accordingly (see middle row in Fig.~\ref{F:9_2_fastslow_appearance}).
This means the total lifetime of slow-appearing CBPs is on average shorter than for fast-appearing ones.
The number of CBPs with short lifetimes (below 6 hr) remains constant with respect to fast or slow appearance.

In the lower row of Fig.~\ref{F:9_2_fastslow_appearance} we see that fast-appearing CBPs have a larger fraction of complex shapes. For the slow-appearing CBPs the number of complex shapes is reduced and the loop-like shapes is increased accordingly.

\subsection{Disappearance}\label{S-disappearance}

\begin{figure}[!htbp]
\centering
\includegraphics[width=8cm]{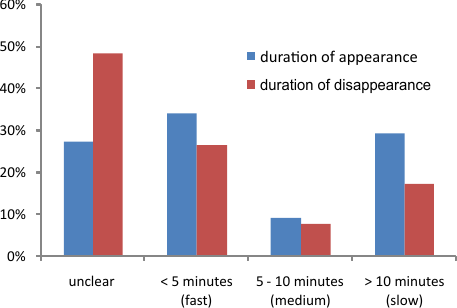}
\caption{Duration of the EUV intensity onset phase (blue) and duration of the CBP disappearance phase (red) (see Sects.~\ref{S-appearance} and ~\ref{S-disappearance}).}
\label{F:fastslow_appearance_a_disappearance_complete}
\end{figure}

In the same way as for the appearance of CBPs we now analyze the duration of their disappearance.
Here, we can use only about 52\% of all CBPs, where the duration is defined (see  Sect.~\ref{S-dataselection}).
We find that fast disappearance of a CBP (92 cases) is about 53\% more frequent than slow disappearance (60 cases) (see Fig.~\ref{F:fastslow_appearance_a_disappearance_complete}).
We also find here a lack of the medium disappearance duration, such as for the appearance.
This means we find a bi-modal distribution of mainly fast- and slow-disappearing CBPs.

In Table~\ref{T:Table_3_9_3_1_Cancellation_Phase_properties} we show different categories of the disappearance of CBPs, where multiple options are possible for the same CBP.
For 69\% of our CBPs the EUV intensity simply fades out.
About 30\% of the disappearing CBPs shrink in size.
We find that 28\% of all CBPs finally merge with another CBP.

\begin{table}[!htbp]
\centering
\caption{CBP disappearance options, multiple choices possible.}
\label{T:Table_3_9_3_1_Cancellation_Phase_properties}
\begin{tabular}{lr}
\hline \hline
Properties of disappearance & Probability\\
\hline
CBP fully disappears and a new CBP follows      &   6.1\% \\
merging with another CBP                                &  28.0\% \\
vanishing intensity                                         &  68.8\% \\
fluctuating intensity                                   &  60.4\% \\
area shrinks                                                &  29.8\% \\
area grows                                                      &   --- \\
shape changes                                               &   5.5\% \\
unclear                                                     &   0.3\% \\
\hline
\end{tabular}
\end{table}

\begin{figure}[!htbp]
\centering
\includegraphics[width=7cm]{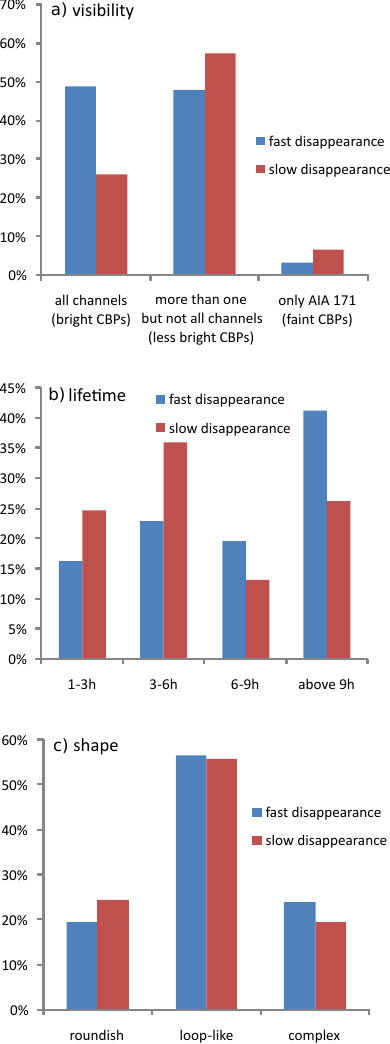}
\caption{Differences between fast (blue) and slow (red) disappearance of CBPs, their visibility in the AIA channels (upper panel), the lifetimes (middle panel), and the shape (lower panel) (see Sect.~\ref{S-disappearance}). The statistics for fast and slow disappearance show the same trend as in Fig.~\ref{F:9_2_fastslow_appearance} panel c).}
\label{F:11_1_fastslow_disappearance_vs_temperature_lifetime}
\end{figure}

Now we separate between fast and slow disappearance and check each group's visibility in EUV and the CBP's lifetime.
The visibility in each AIA channel varies for brighter and fainter CBPs.
In panel a) of Fig.~\ref{F:11_1_fastslow_disappearance_vs_temperature_lifetime} we show that the groups of brighter and less bright CBPs have roughly an equal share for the group of fast disappearing CBPs.
We find that the group of slow-disappearing CBPs is less often visible in all AIA channels.
At the same time, the share of less bright CBPs is significantly enhanced for slow-disappearing CBPs.

In panel b) of Fig.~\ref{F:11_1_fastslow_disappearance_vs_temperature_lifetime} we see that 61\% of the fast-disappearing CBPs have lifetimes of 6 hr or longer.
For slow-disappearing CBPs we find exactly the opposite behavior, where 61\% of them have lifetimes of less than 6 hr.

There is no significant difference between the shapes of CBPs depending on their fast or slow disappearance (see panel c) of Fig.~\ref{F:11_1_fastslow_disappearance_vs_temperature_lifetime}).
This distribution is similar to that for the appearance of CBPs (see lower row in Fig.~\ref{F:9_2_fastslow_appearance}).

\subsection{Lifetime}\label{S-lifetime}

\begin{figure}[!htbp]
\centering
\includegraphics[width=8.8cm]{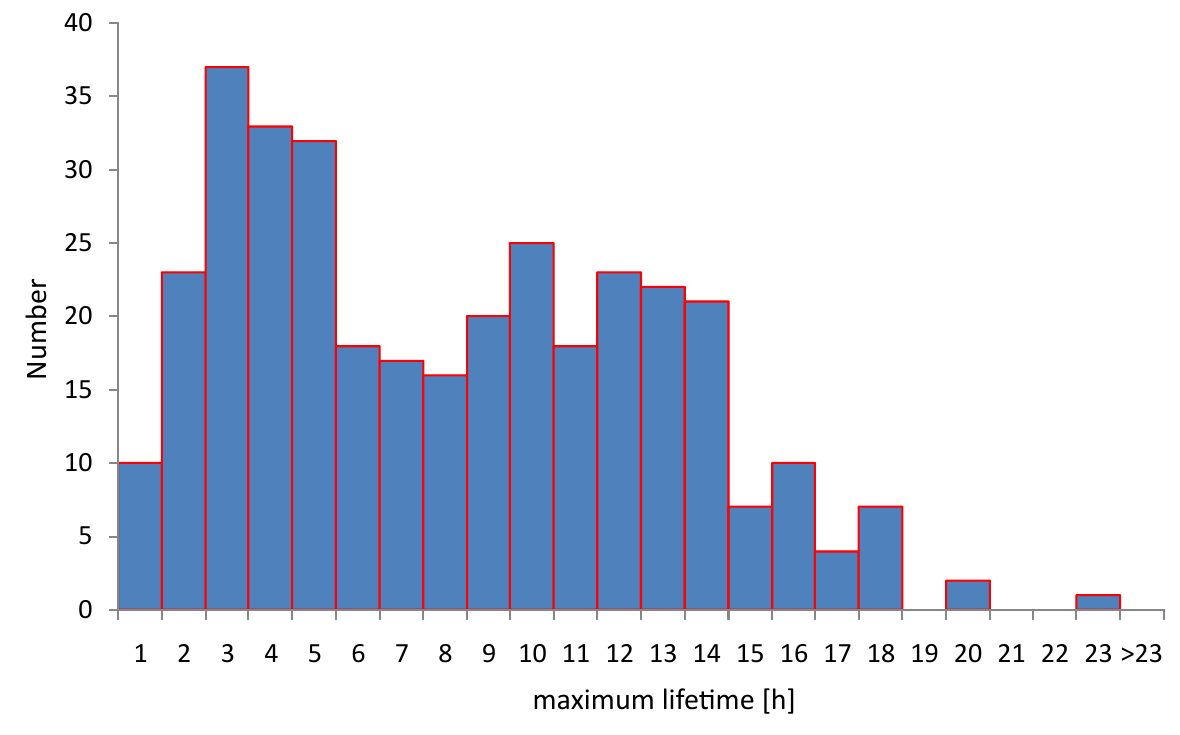}
\caption{Lifetimes of CBPs that are fully tracked from appearance to disappearance (see Sect.~\ref{S-lifetime}).}
\label{F:3_Lifetime}
\end{figure}

We now investigate the total lifetime of CBPs (see Fig.~\ref{F:3_Lifetime}).
We find that about 20\% of CBPs have a lifetime shorter than 3 hr.
The interval with lifetimes of 3--6 hr contains the largest fraction of CBPs.
More than half of all CBPs have a lifetime of less than 9 hr.
Only one CBP reaches a lifetime of 21--24 hr, and no CBP in our sample lives for longer than 24 hr.

\subsection{Visibility versus shape}\label{S-Visibility-shape}

The group of ``bright'' CBPs is defined as those visible in all AIA channels, while ``less bright'' CBPs are visible in more than one but not all channels and do not include the ``faint'' CBPs that are visible only in the AIA 171 channel (see Sect.~\ref{S-visibility}).
We find that bright CBPs have a slightly higher probability (42\% instead of the average 36\%) to merge with another CBP, while this probability is 30\% for less bright CBPs that are not visible in all AIA channels.
In contrast, we do not find any merging faint CBPs.

\begin{figure}[!htbp]
\centering
\includegraphics[width=7cm]{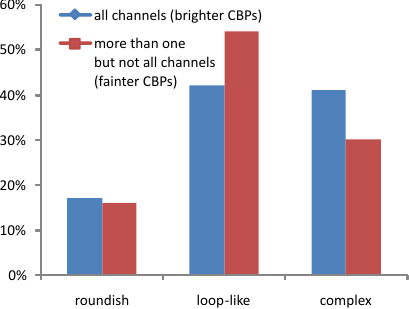}
\caption{Distribution of shapes for brighter CBPs that are visible in all AIA channels (blue) vs. fainter CBPs that are not visible in all channels (red) (see Sect.~\ref{S-Visibility-shape}).}
\label{F:Shape_Hot_vs_faint_CBP}
\end{figure}

There are significantly more CBPs with a complex shape and good visibility in all AIA channels (41\%) (see Fig.~\ref{F:Shape_Hot_vs_faint_CBP}).
Less bright CBPs (not visible in all channels) have significantly less complex shapes (30\%) than the bright ones.
Likewise, we find less round shapes with high brightness in all channels (17\%) and similar for less bright CBPs (16\%).
The loop-like structures are more common for less bright CBPs (see Fig.~\ref{F:Shape_Hot_vs_faint_CBP}).
Overall averages of the different shapes can be found in Sect.~\ref{S-shape}.

\subsection{Visibility versus lifetime}\label{S-visibilityvslifetime}

\begin{figure}[!htbp]
\centering
\includegraphics[width=7cm]{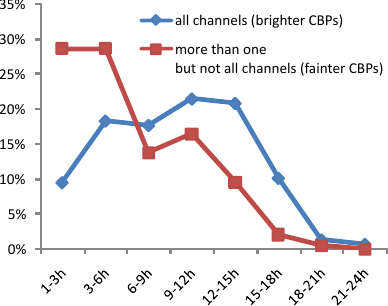}
\caption{Lifetimes of brighter CBPs visible in all AIA channels (blue) vs. fainter CBPs not visible in all channels (red) (see Sect.~\ref{S-visibilityvslifetime}).}
\label{F:Lifetime_Hot_vs_faint_CBP}
\end{figure}

On one hand, we find that only about 28\% of the bright CBPs (visible in all channels) have a lifetime of less than 6 hr (see Fig.~\ref{F:Lifetime_Hot_vs_faint_CBP}), while this is the case for 44\% for the global average, see Sect.~\ref{S-lifetime}.
On the other hand, more than 57\% of the fainter CBPs (not visible in all channels) exist for 6 hr or less.
This indicates a statistically significant trend, where the lifetime of CBPs depends on their EUV brightness.
In particular, the majority of CBPs that are visible in all AIA channels (brighter ones) remain visible for more than 6 hr.
In constrast, the majority of those CBPs that are not visible in all channels (fainter ones) exist for a maximum of 6 hr.

\subsection{Lifetime versus shape}\label{S-lifetimevsshape}

\begin{figure}[!htbp]
\centering
\includegraphics[width=8.8cm]{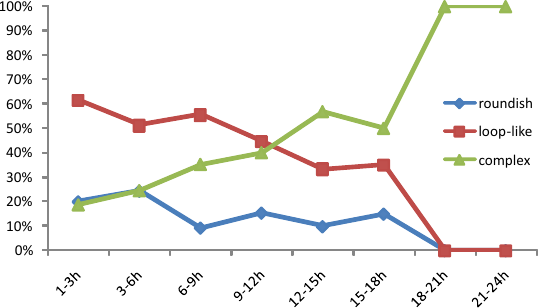}
\caption{Fraction of the observed CBP shapes in three-hour intervals of the lifetime, each bin is normalized to 100\%.
See Figure~\ref{F:3_Lifetime} for an absolute distribution of the CBP lifetimes; also see Sect.~\ref{S-lifetime}.}
\label{F:1_Lifetime_vs_Shape}
\end{figure}

In Section~\ref{S-shape} we show the fraction of the shape categories of all CBPs.
Figure~\ref{F:1_Lifetime_vs_Shape} displays the fraction of each shape category depending on CBP lifetime intervals.
We find a general trend that roundish and loop-like shapes become less frequent with longer lifetimes of the CBPs.
The turnover point seems to be at a lifetime of 12 hr or more, where complex shapes are then 50\% or more.

\section{Conclusions}

A typical CBP lifetime is about 3--6 hr.
We find that most of the CBPs are visible in all AIA wavelength channels, which suggests a high temperature above one million Kelvin.
About one-third of the CBPs merge for a short time with another CBP.
We find no influence of mergers on the growing behavior of CBPs.
The most frequent shape of CBPs is loop-like, followed by complex; the least frequent shape is round.

Even though our observed CBP sample is much smaller than \cite{2015ApJ...807..175A} use, their results are in good agreement with ours if we only consider their CBPs that live longer than 30 minutes.
We do not consider phenomena that live less than 30 minutes to be CBPs; instead, they are probably tracking failures.

From Figure~\ref{F:Figure_3_3_BP_intensities_images45_bp165} we can see that all peaks in the EUV emission appear at the same time in all AIA channels.
This means that we do not see a slowly cooling plasma packet, nor do we see a hot plasma transport mechanism to the corona.
Instead, this emission is produced by the same coronal plasma packet at the same time, which means the heating must happen locally in the corona.

It would be interesting to distinguish between different magnetic regions, such as bipolar, multipolar, or mostly quiet-Sun.
We propose to ask this question in a follow-up study on the magnetic foot points of CBPs.
This would give an insight into whether certain shapes occur mostly from loop-like magnetic fields.
We expect that complex CBPs occur from multipolar foot points in the photosphere.
Regarding this work, we do not see significant differences in the coronal co-rotation behavior due to the magnetic configuration below the investigated CBPs.

In our whole CBP sample set we find no cases of a height-dependent differential rotation between the photosphere and the lower corona.
This means that the lower corona still rotates differentially depending on the latitude, but exactly following the same differential rotation as seen in the photosphere.
This can also be seen in our online videos, where the magnetic features observed in HMI in the photosphere remain over a long time at the same position as the EUV-bright features observed in AIA. Without this co-rotation one would see a systematic drift in the video, which is not there.
When we tried out other coronal differential rotation profiles, we found the tracking of the CBPs to be unstable.
This makes sense if we assume there is a strong magnetic connectivity between the lower corona and the photosphere, 
especially for CBPs, where we usually see stronger magnetic fields than in the quiet-Sun, leading to a plasma beta that is particularly lower than for the coronal plasma above quiet-Sun \citep{Bourdin:2017_beta}.
Hence, the coronal plasma is more tightly connected to mostly loop-like field lines that are rooted at the CBP foot-point polarities in the photosphere.
We conclude that the lower solar atmosphere rotates as a rigid body together with the photosphere.

\begin{acknowledgements}
SDO Data supplied courtesy of the SDO/HMI and SDO/AIA consortia. This research was partly funded by a grant of the Austrian Science Fund (FWF): P~32958-N.    
\end{acknowledgements}

\newpage

\bibliography{Literatur_Isabella}
\bibliographystyle{aa}

\end{document}